\documentclass[]{spie}  
\usepackage{todonotes}

\usepackage{amsmath,amsfonts,amssymb}
\usepackage{booktabs}
\usepackage{float}
\usepackage{graphicx}
\usepackage{gensymb}
\usepackage[colorlinks=true, allcolors=blue]{hyperref}
\usepackage{subcaption}

\usepackage{lmodern}
\usepackage{siunitx}
\usepackage{comment}
\usepackage[justification=centering]{caption}

\title{MOSAIC at ELT:  Design and First Prototyping of Novel Robotic Optical-Relay Positioners}

\author[a]{Maxime Rombach}
\author[b]{Markus Thurneysen}
\author[a]{Lucas Ortolani}
\author[c]{Jurgen Schmoll}
\author[a]{Diane Chapuis}
\author[a]{Malak Galal}
\author[a]{Sebastien Pernecker}
\author[e]{Alec York}
\author[e]{Ellen Schallig}
\author[e]{Ian Lewis}
\author[e]{Louis Gabarra}
\author[d]{Cassio Berni}
\author[d]{Ojonugwa Adukwu}
\author[f]{Gavin Dalton}
\author[d]{Fabio Fialho}
\author[a]{Michaela Hirschmann}
\author[a]{Jean-Paul Kneib}
\author[g]{Thomas Augusteijn}
\author[h]{Kacem El Hadi}
\author[i]{Marisa Garcia-Vargas}
\author[j]{Marie Larrieu}
\author[k]{Alessandro Meoli}
\author[c]{Tim Morris}
\author[h]{Roser Pello}
\author[l]{Mathieu Puech}
\author[m]{Myriam Rodrigues}
\author[n]{Rubén Sanchez-Janssen}
\author[m]{Alondra Sola}
\author[o]{Jay Stephan}
\author[h]{Christian Surace}

\affil[a]{Institute of Physics, Laboratory of Astrophysics, Ecole Polytechnique Federale de Lausanne (EPFL), Observatoire de Sauverny, CH-1290 Versoix, Switzerland}
\affil[b]{Haute Ecole du Paysage, d'Ingénierie et d'Architecture de Genève (HEPIA), Hautes Ecoles Spécialisées de Suisse Orientale (HES-SO), Rue de la prairie 4, 1202 Geneva, Switzerland}
\affil[c]{Centre for Advanced Instrumentation, Department of Physics, Durham University, South Road, Durham DH1 3LE, UK}
\affil[d]{University of Sao Paulo, Polytechnic School, Department of Electrical Engineering, Sao Paulo, Brazil}
\affil[e]{Department of Physics, University of Oxford, Keble Road, Oxford, OX13RH, UK}
\affil[f]{STFC RALSpace, Rutherford Appleton Laboratory, Didcot, OX11 0QX, UK}
\affil[g]{NOVA Optical IR Instrumentation Group, ASTRON, Netherlands}
\affil[h]{Aix Marseille Université, CNRS, CNES, LAM (Laboratoire d’Astrophysique de Marseille), UMR 7326, 13388 Marseille, France}
\affil[i]{FRACTAL SLNE, Calle Tulipán 2, portal 13, 1A, E-28231 Las Rozas de Madrid, Spain}
\affil[j]{Univ Toulouse, CNES, CNRS, IRAP, F-31400 Toulouse, France}
\affil[k]{European Southern Observatory, Garching, Germany}
\affil[l]{Laboratoire d’étude de l’Univers et des phénomènes eXtrêmes (LUX), Observatoire de Paris, PSL University, CNRS, 92190 Meudon, France
UNIDIA, Observatoire de Paris, PSL University, CNRS, 92190 Meudon, France}
\affil[m]{UNIDIA, Observatoire de Paris, PSL University, CNRS, 92190 Meudon, France}
\affil[n]{Isaac Newton Group of Telescopes, Santa Cruz de La Palma, Spain}
\affil[o]{STFC UK Astronomy Technology Centre, Royal Observatory, Blackford Hill, Edinburgh EH9
3HJ, UK}

\authorinfo{Further author information: (Send correspondence to M. R.)\\M.R.: E-mail: maxime.rombach@epfl.ch, Telephone: +41 (0) 21 693 28 76}

\begin{document} 
\maketitle
\thispagestyle{empty}

\begin{abstract}
The Extremely Large Telescope (ELT) is, to date, the most ambitious ground-based telescope under construction. MOSAIC is a multi-objects spectrograph (MOS) that aims to make full use of the largest telescope in the world. At its heart, about 250 robotic positioners will pick-off skylight from the focal surface of the ELT to feed it to its Near Infrared (NIR) and Visible (VIS) spectrographs.
The gigantic scale of the ELT presents three main challenges for MOSAIC positioners: (1) the light beams on the focal surface cannot be focused in a single fiber, similarly to other MOS instruments, involving a design with relay mirrors patrolling the field of view, and re-imaging the sub-field on 2 fixed fiber bundles located 600 mm behind the ELT focal plane (2) The positioner needs to adapt to the local telecentricity, which means it has to point at the ELT pupil center located 37.868 m away from the focal plane  (3) The Atmospheric Dispersion Corrector (ADC) needed to cover the whole focal surface of the ELT is impossible to build to this scale; hence each positioner needs its own ADC.
EPFL is responsible for designing and supervising the mass manufacturing of the positioners. This paper aims to present its initial design and prototypes.
\end{abstract}


\section{INTRODUCTION}
\label{sec:intro}  
\subsection{Overview of the instrument}
MOSAIC (Multi-Object Spectrograph for Astrophysics, Intergalactic-medium studies and Cosmology) is part of the second suite of instruments for the ELT. It is located on one of the two nasmyth platforms of the ELT, right in the optical axis of the Prefocal Station.\\

\begin{figure}[H]
    \centering
    \captionsetup{width=.7\linewidth}
    \includegraphics[width=0.5\linewidth]{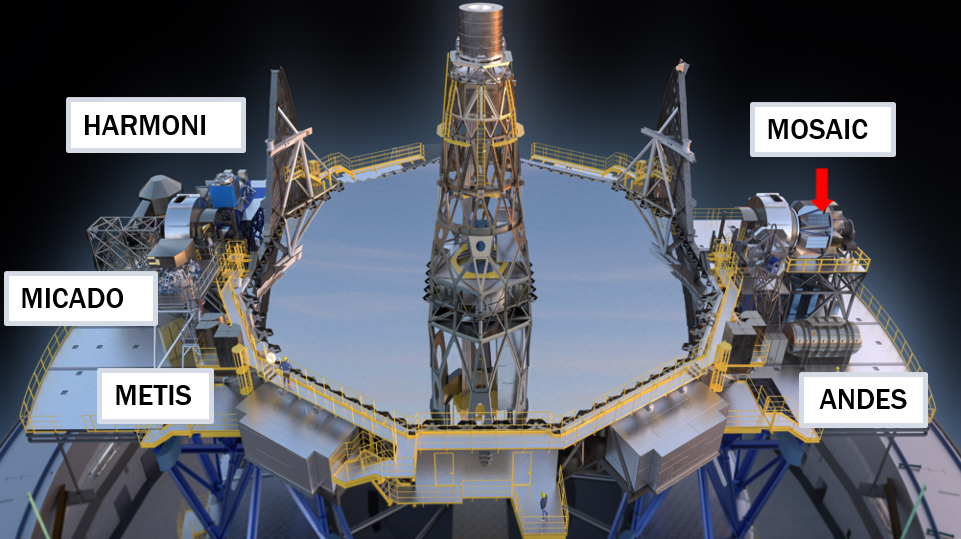}
         \vspace{0.2cm}
    \caption{Artist’s impression of the ELT instruments (2021); Instruments positions may have changed since. Credit: ESO}
    \label{fig:elt_instruments}
\end{figure}

\begin{figure}[H]
     \centering
     \begin{subfigure}[b]{0.49\textwidth}
         \centering
         \includegraphics[width=0.7\textwidth]{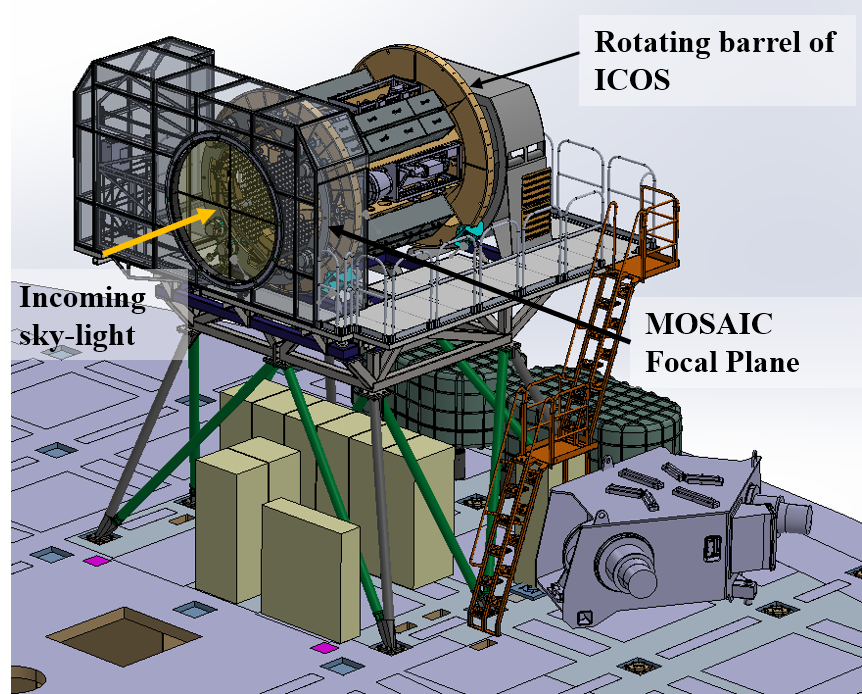}
         \caption{CAD view of the preliminary design of the instrument; highlighting the incoming sky-light direction and the rotating barrel location.\\
         Credit: MOSAIC consortium}
         \label{fig:icos}
     \end{subfigure}
     \hfill
     \begin{subfigure}[b]{0.49\textwidth}
         \centering
         \includegraphics[width=\textwidth]{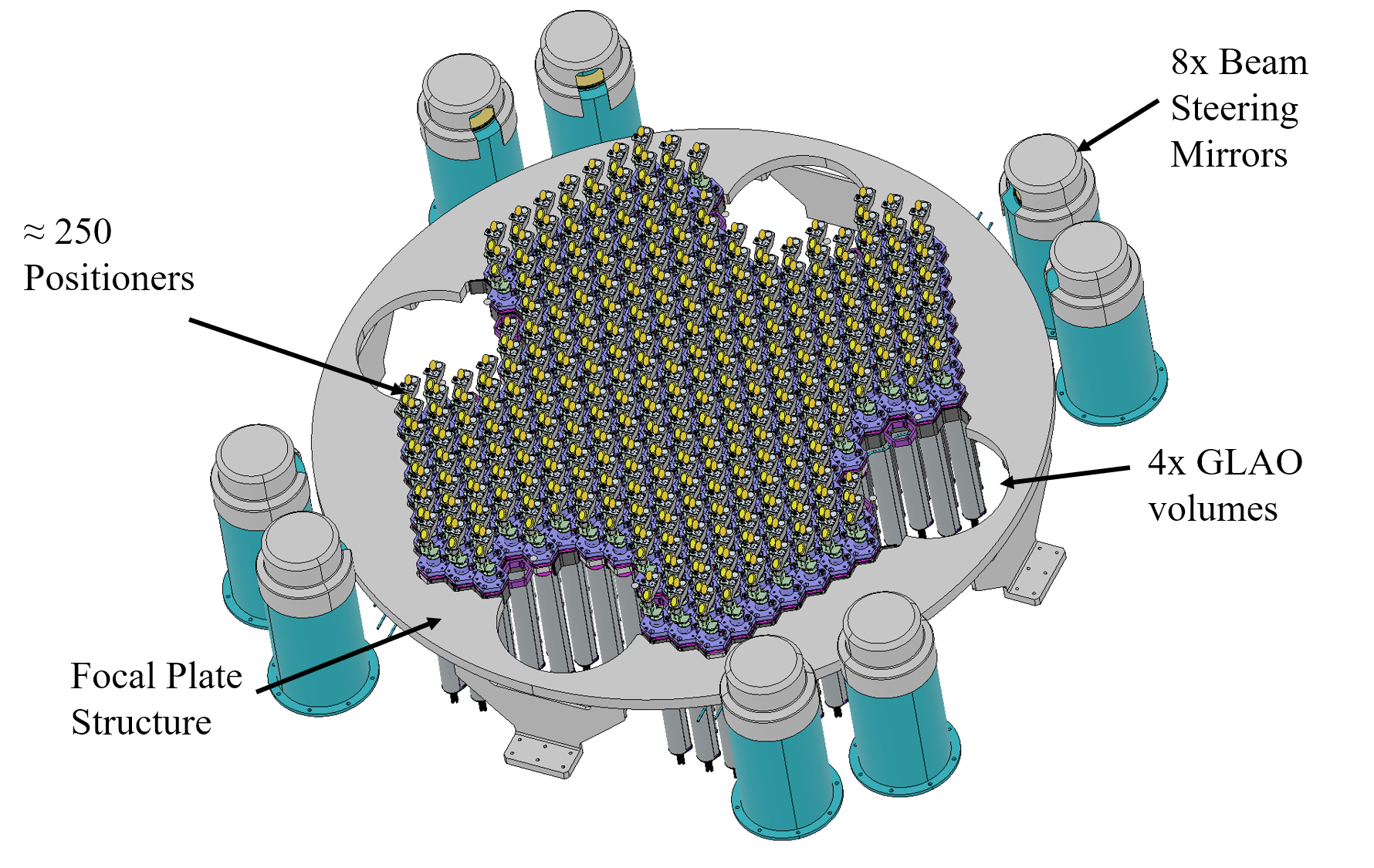}
         \caption{CAD view of the preliminary design of the MOSAIC Focal Plane assembly. \\ Credit: Oxford University}
         \label{fig:fp}
     \end{subfigure}
        \vspace{0.2cm}
        \caption{Preliminary CAD views of the overall location of the positioners}
        \label{fig:icos&fp}
\end{figure}

The positioners are part of the MOSAIC \textit{Focal Plane assembly}\cite{schallig_mosaic_2026}, under the responsability of Oxford University (Figure \ref{fig:fp}). The latter is mounted on a rotating barrel oriented 90° to the vertical axis (Figure \ref{fig:icos}) in a similar fashion as MOONS\cite{gonzalez_moons_2022} for VLT.

\noindent This paper aims to give an overview of the preliminary design of the MOSAIC positioner and early prototyping.

\subsection{POS modes}

\begin{figure}[H]
    \centering
    \captionsetup{width=.5\linewidth}
    \includegraphics[width=0.8\linewidth]{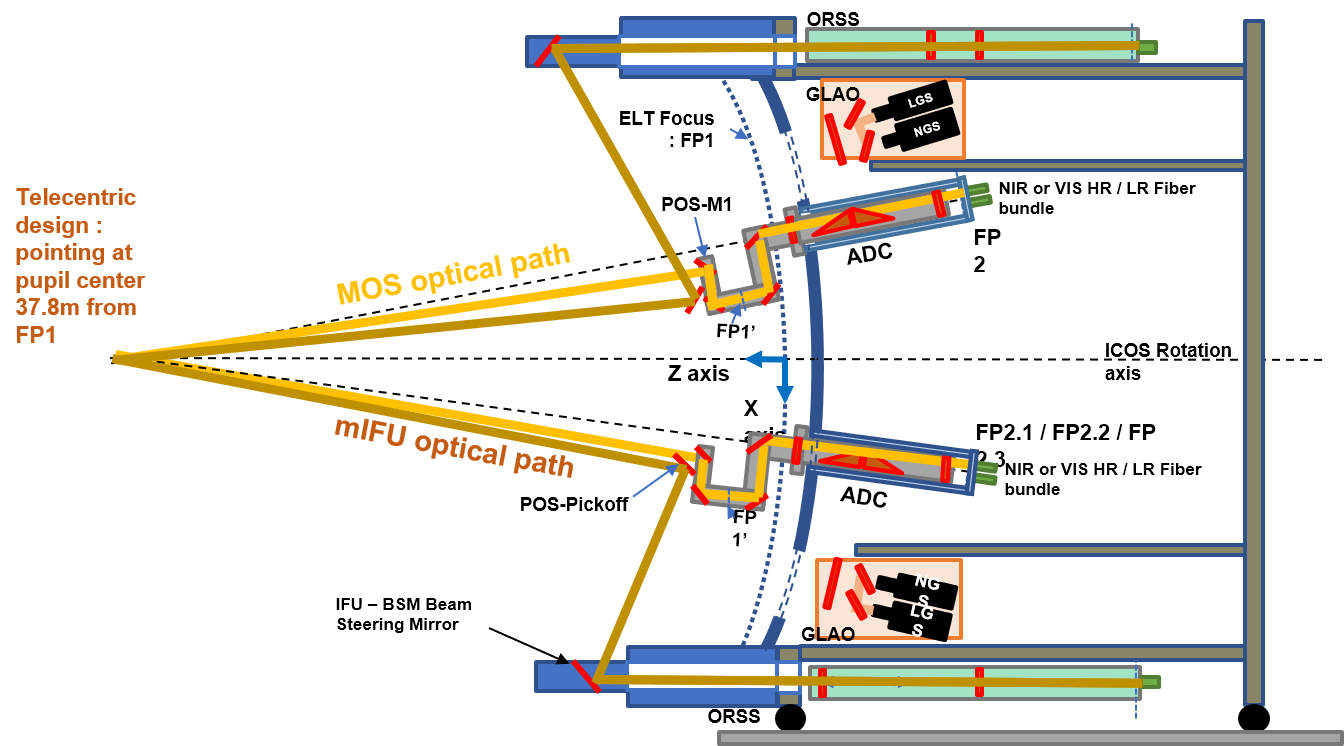}
    \vspace{0.2cm}
    \caption{MOSAIC Front-End Optical Model schemtics. Credit: Diane Chapuis - EPFL}
    \label{fig:pos_modes}
\end{figure}

MOSAIC is born from the fusion of two instrument concepts OPTIMOS-EVE\cite{hammer_optimos-eve_2010} and EAGLE\cite{morris_eagle_2012}. It aims to combine MOS and Integral Field Unit (IFU) capabilities for a wavelength range of 390 nm to 1800 nm. The positioners are the core elements that will enable MOSAIC to switch from one mode to the other via is two different optical paths as seen in Figure \ref{fig:pos_modes}. One positioner can be used for one or the other but \textit{not both at the same time}:

\begin{itemize}
    \item \textbf{MOS mode:} the positioner needs to place the entrance of the \textit{MOS optical path} at the sky target location. Light will travel through the four relay mirrors, be collimated by the first lens L1 for the Atmospheric Dispersion Corrector (ADC) prisms to correct the aberration from the atmosphere and focus light in either the Near InfraRed (NIR) or Visible (VIS) fiber bundle; see light yellow path in Figure \ref{fig:pos_modes}
    \item \textbf{mIFU mode:} the positioner places the pickoff mirror located at its top at the target location. Light will travel towards one of the eight Beam Steering Mirror (BSM) of the Focal Plane, to be injected in the Optical Relay SubSystem (ORSS)\cite{cvetojevic_mosaic_2026}; see dark yellow path in Figure \ref{fig:pos_modes}.
\end{itemize}

 The presented work will focus on the mechanical implementation of the MOSAIC positioner. Its optical design is detailed in Schmoll, et al.(2026)\cite{schmoll_mosaic_2026}.
\section{POS design}
\subsection{POS overview}
\begin{figure}[H]
    \centering
    \includegraphics[width=0.7\linewidth]{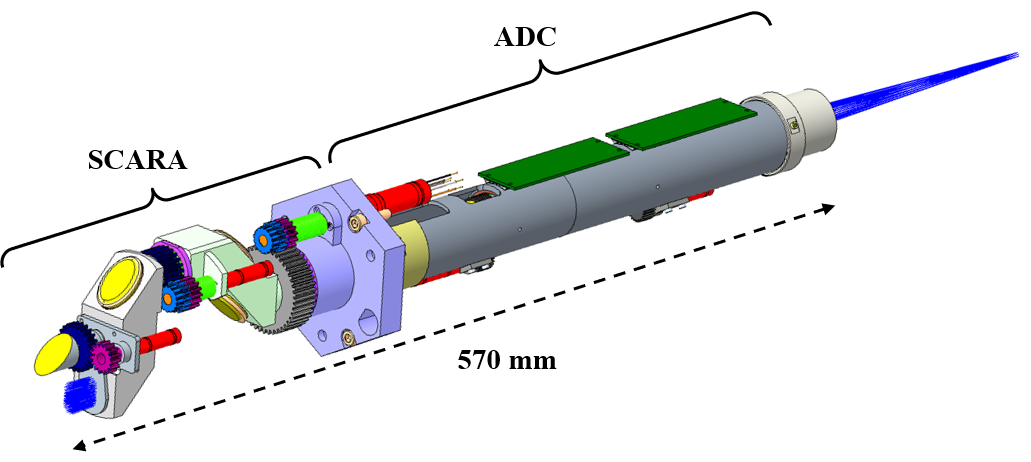}
    \vspace{0.2cm}
    \caption{3D model of the preliminary design of the MOSAIC positioner: POS SCARA and POS ADC}
    \label{fig:iso_view}
\end{figure}

The MOSAIC positioner is decoupled into two sub-assemblies namely: POS SCARA and POS ADC. They are bolted together on the positioner's hexagonal base, in light purple in Figure \ref{fig:iso_view}. EPFL is responsible in the design, prototyping and supplying of the $\approx$ 250 robots that will be assembled in the Focal Plane by Oxford University. Once the design and prototyping phase are complete we will turn to manufacturers to industrialize the positioner and manufacture the series.

\begin{figure}[H]
    \centering
    \includegraphics[width=0.9\linewidth]{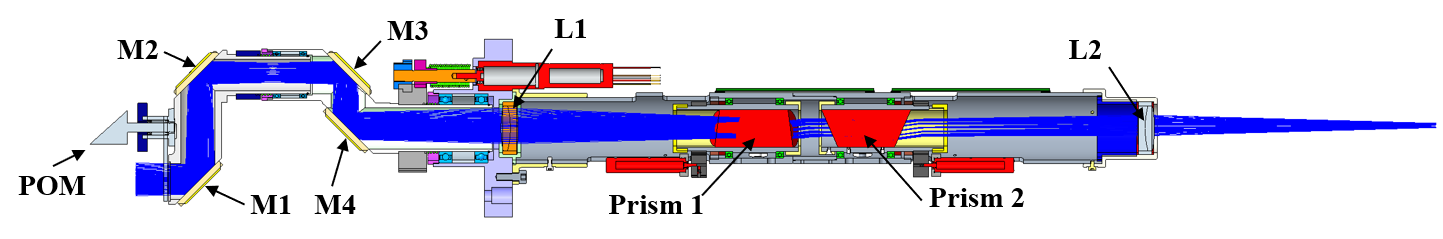}
    \vspace{0.2cm}
    \caption{Cut view of the positioner highlighting the optical elements}
    \label{fig:cut_view}
\end{figure}

The mechanics of the MOSAIC positioner is largely built around the MOS path optical train which needs to accommodate MOSAIC's wide wavelength range; 390 nm to 1800 nm. The context of the optical train is only reminded here, one can find more details in Schmoll, et al.(2026)\cite{schmoll_mosaic_2026}. It is composed of the following elements that one can visualize in Figure \ref{fig:cut_view}:
\begin{itemize}
    \item 4 mirrors (M1 - M4): identical relay mirrors actuated by the alpha and beta arms to pick-off light from the ELT focal plane
    \item 2 lenses (L1-L2): identical lenses mounted symmetrically to collimate sky-light at L1 and refocus it in the desired fiber bundle at L2
    \item 2 prisms (Prism 1 and Prism 2): identical triplet prisms that are in charge of correcting chromatic aberrations from the atmosphere before focusing light in fiber bundles
    \item 1 Pickoff mirror (POM): located at the top of the positioner, this mirror can steer a light beam from the ELT focal plane towards the mIFU path
\end{itemize}
\subsection{POS SCARA}
\subsubsection{POS Patrol Area}
\begin{figure}[H]
     \centering
     \begin{subfigure}[b]{0.49\textwidth}
         \centering
         \includegraphics[width=0.6\textwidth]{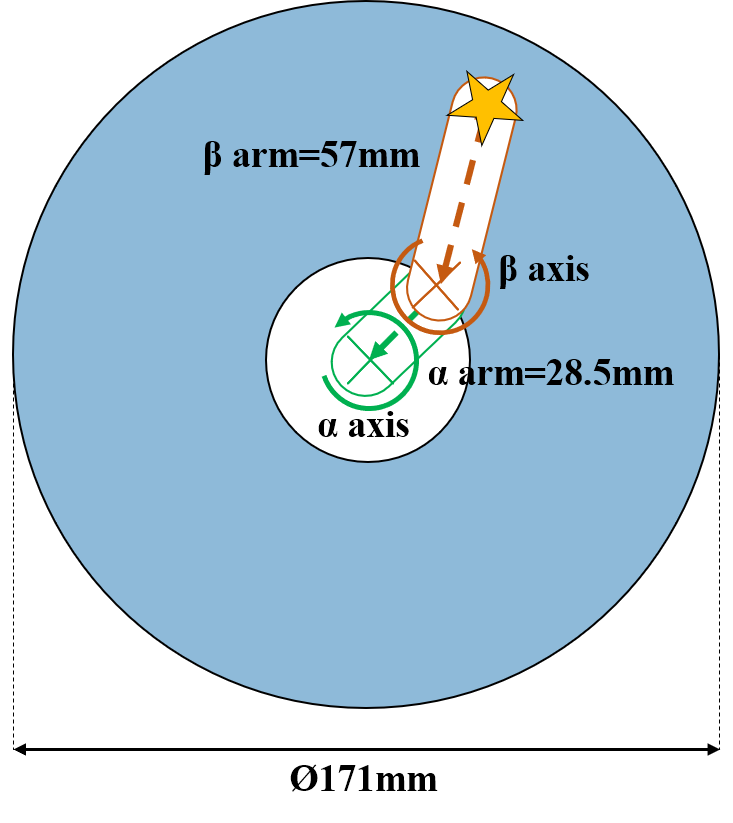}
         \caption{Schematics of the arms dimensions forming the positioner's workspace: its patrol radius is 85.5 mm, the middle blind area has a radius of 28.5 mm }
         \label{fig:sec2_workspace}
     \end{subfigure}
     \hfill
     \begin{subfigure}[b]{0.49\textwidth}
         \centering
         \includegraphics[width=0.6\textwidth]{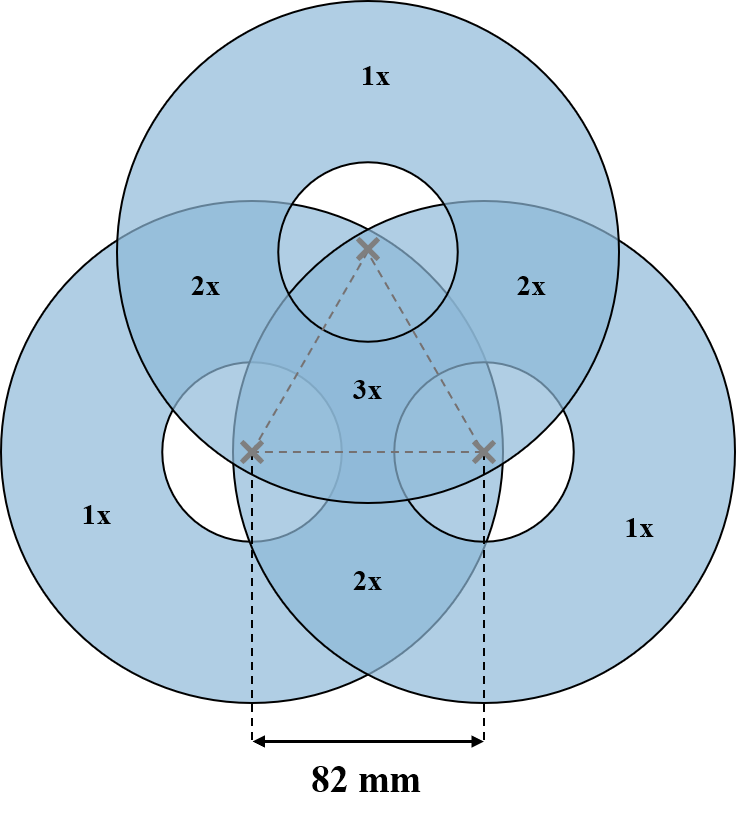}
         \caption{Neighboring positioners patrol area overlap with a \textit{pitch} of 82 mm; highlighting regions reachable by several positioners}
         \label{fig:sec2_overlap}
     \end{subfigure}
     \vspace{0.2cm}
    \caption{Schematics of the MOSAIC positioner patrol area}
    \label{fig:sec2_workspace_overlap}
\end{figure}

The POS SCARA part is the sub-assembly enabling the positioner to pick light from its large workspace in the ELT focal plane. Coming from the industrial robot domain, the Selective Compliance Articulated Robot Arm (SCARA) architecture is also known as \textit{theta-phi} or \textit{alpha-beta} configuration and is already used in successful instruments such as SDSS-V\cite{sayres_sdss-v_2022}, DESI\cite{schubnell_desi_2016} or MOONS\cite{gonzalez_moons_2022}. Its workspace, or patrol area, is defined by the distance between M1-M2 and M3-M4 center. The first one is later referred as the \textit{beta arm length} and the latter \textit{alpha arm length}. Figure \ref{fig:sec2_workspace_overlap} gives an overview of the kinematics of the positioner, where the blue regions are the spaces where the positioners can pick-off light form the focal plane.

\subsubsection{The ELT non-telecentricity challenge}

\begin{figure}[H]
    \centering
    \includegraphics[width=0.55\linewidth]{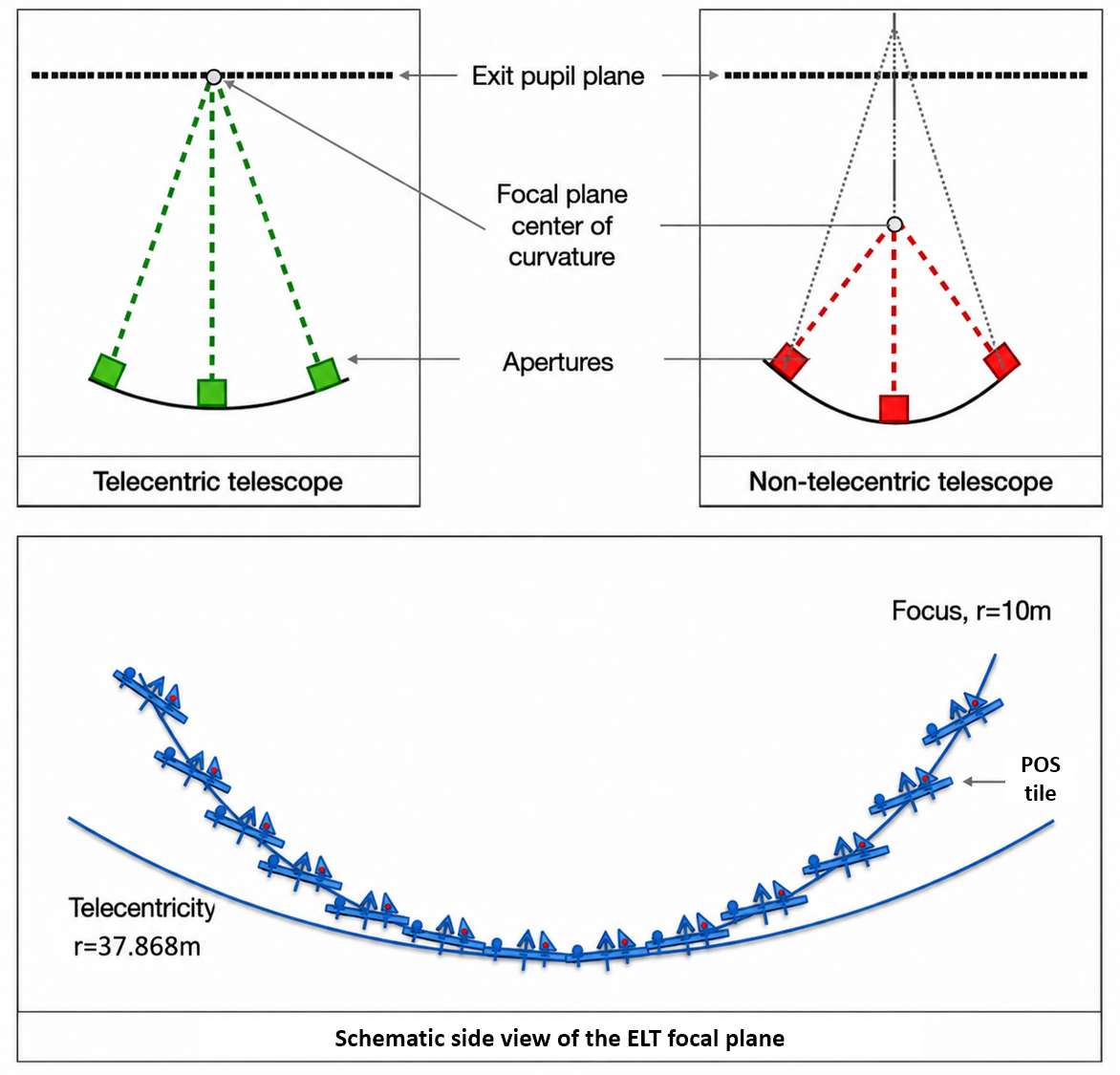}
    \vspace{0.2cm}
    \caption{(Top left) Ideal case of a telecentric telescope where aligning apertures with the curved focal surface is sufficient to point at the pupil. (Top right) ELT non-telecentric case: focal plane curvature center and exit pupil do not match. (Bottom) Proposed positioners tile arrangement for MOSAIC. The tiles are set to the focus curvature but point at the telecentric center}
    \label{fig:sec_2_telecentricity}
\end{figure}

The ELT is a non-telecentric telescope. Consequently the ELT focal plane curvature (r=10 m; see Figure \ref{fig:sec_2_telecentricity}) does not point to the exit pupil plane, 37.868 m away. Figure \ref{fig:sec_2_telecentricity} from Rodrigues, et al. (2016)\cite{rodrigues_developing_2016} shows the trade-off that the instrument is taking. The positioners tiles center coincides with the ELT focal plane while pointing at the telecentric center, 37.868 m away; see bottom image of Figure \ref{fig:sec_2_telecentricity}. The patrol area of the positioners will therefore be on the \textit{average focus}, not always in focus, but will always point towards the telecentric center.

\noindent We can now isolate one tile from Figure \ref{fig:sec_2_telecentricity} and study the effect of the ELT non-telecentricity. In order for the positioner to pick off light from the ELT focal plane, its internal axes of the positioner need to point at the pupil center, located 37.868 m away from the focal surface.

\begin{figure}[H]
    \centering
    \includegraphics[width=0.5\linewidth]{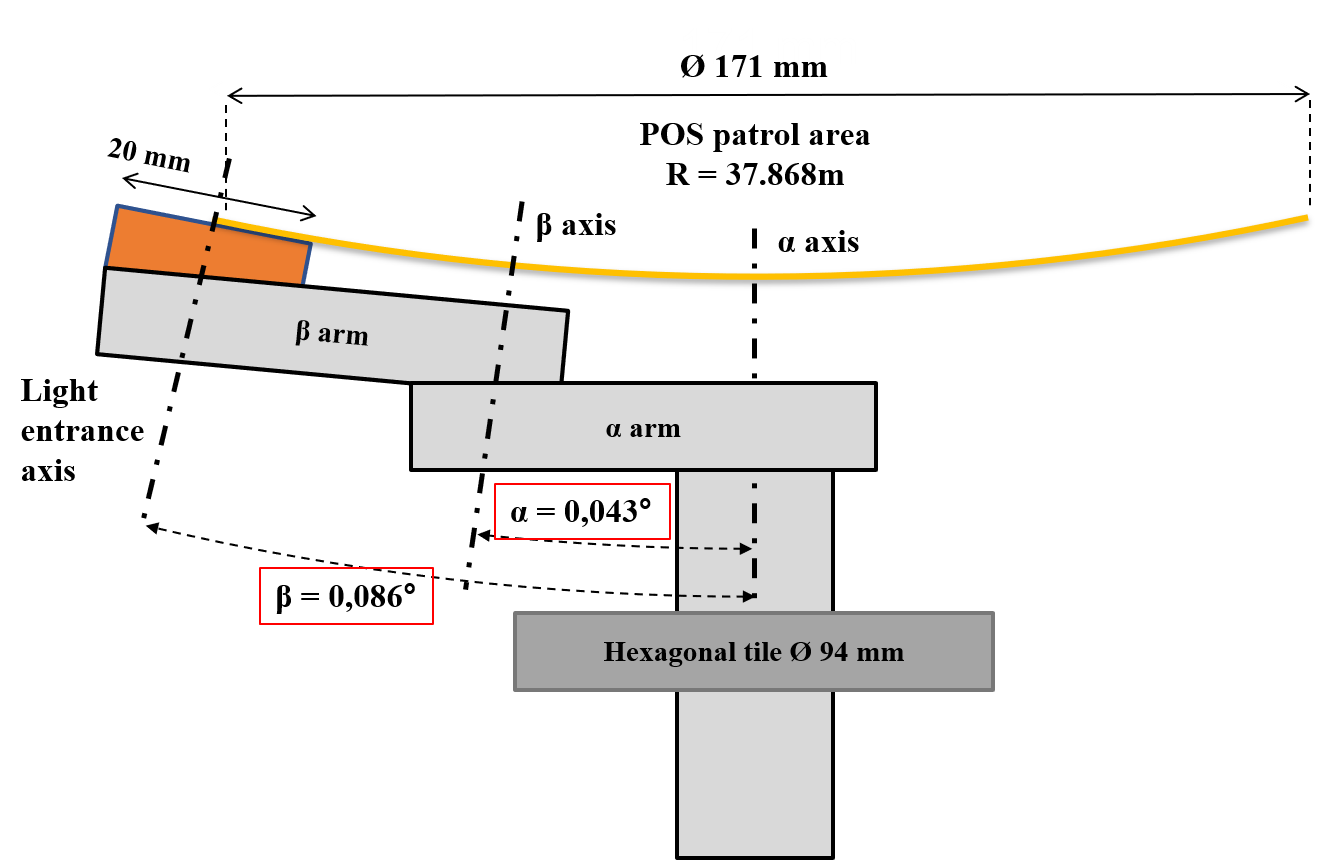}
    \vspace{0.2cm}
    \caption{Schematics of the POS internal axes required angles to always point at the pupil center regardless of the arms positions}
    \label{fig:sec_2_pos_telecentricity}
\end{figure}

\noindent Given the alpha arm length (28.5mm), the beta arm length (57mm) and the radius, R, of curvature pointing to the telescope pupil (37.868m) we can deduce the necessary pointing angles from trigonometry:
\begin{align} \label{eq:pos_telecentricity}
\begin{split}
\alpha &=  \arctan \left(\dfrac{\alpha_\text{arm length}}{R}\right) = 0.043\degree \\ 
\beta &=  \arctan \left(\dfrac{\beta_\text{arm length}}{R}\right) = 0.086\degree
\end{split}
\end{align}

\noindent These pointing angles are going to prove challenging in the manufacturing of the positioners in terms of the parts themselves and also optics alignment. The MOS path mirrors angles will, indeed, be affected by those results as shown in Table \ref{tab:mirrors_faces_angles}.

\subsubsection{POS SCARA mechanics}

\begin{figure}[H]
     \centering
     \begin{subfigure}[b]{0.49\textwidth}
        \centering
        \includegraphics[width=\linewidth]{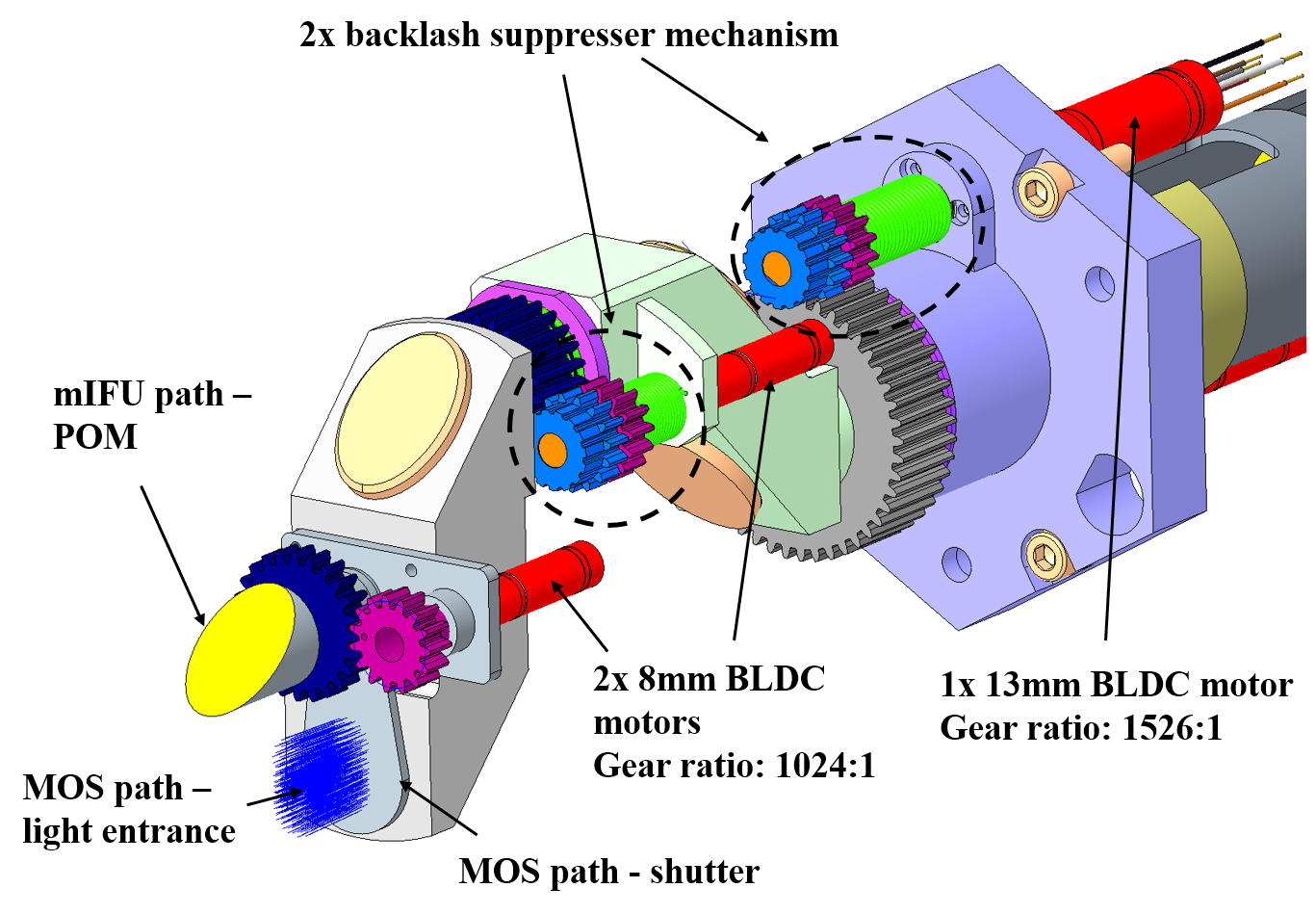}
        \caption{Front side}
        \label{fig:pos_scara_meca_front}
     \end{subfigure}
     \hfill
     \begin{subfigure}[b]{0.49\textwidth}
         \centering
         \includegraphics[width=0.8\textwidth]{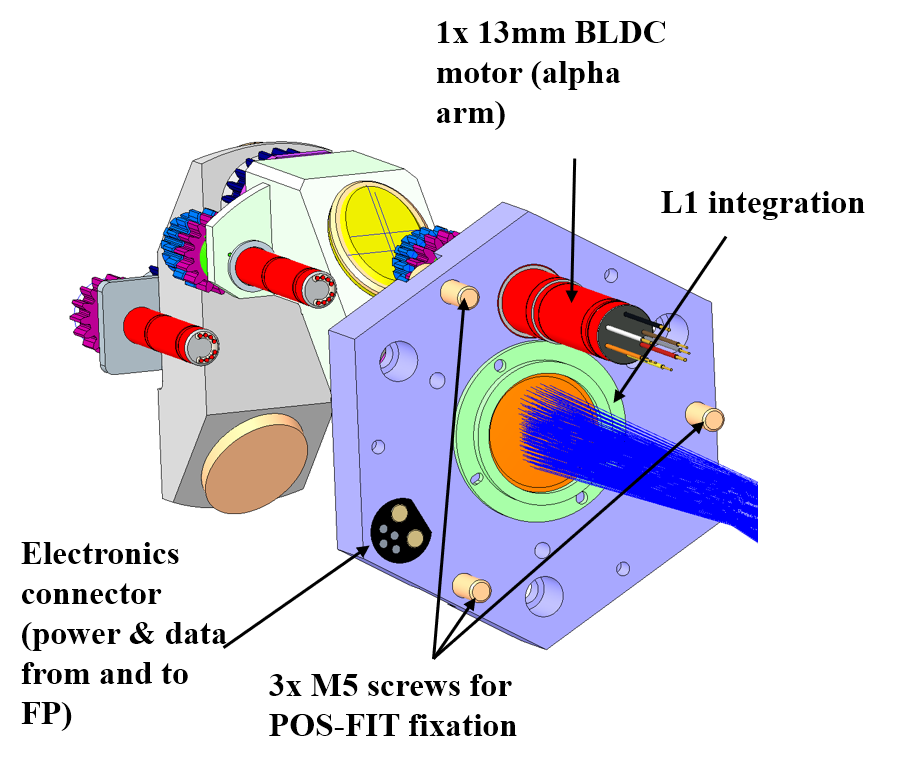}
         \caption{Back side}
         \label{fig:pos_scara_meca_back}
     \end{subfigure}
          \begin{subfigure}[b]{0.7\textwidth}
         \centering
         \includegraphics[width=\textwidth]{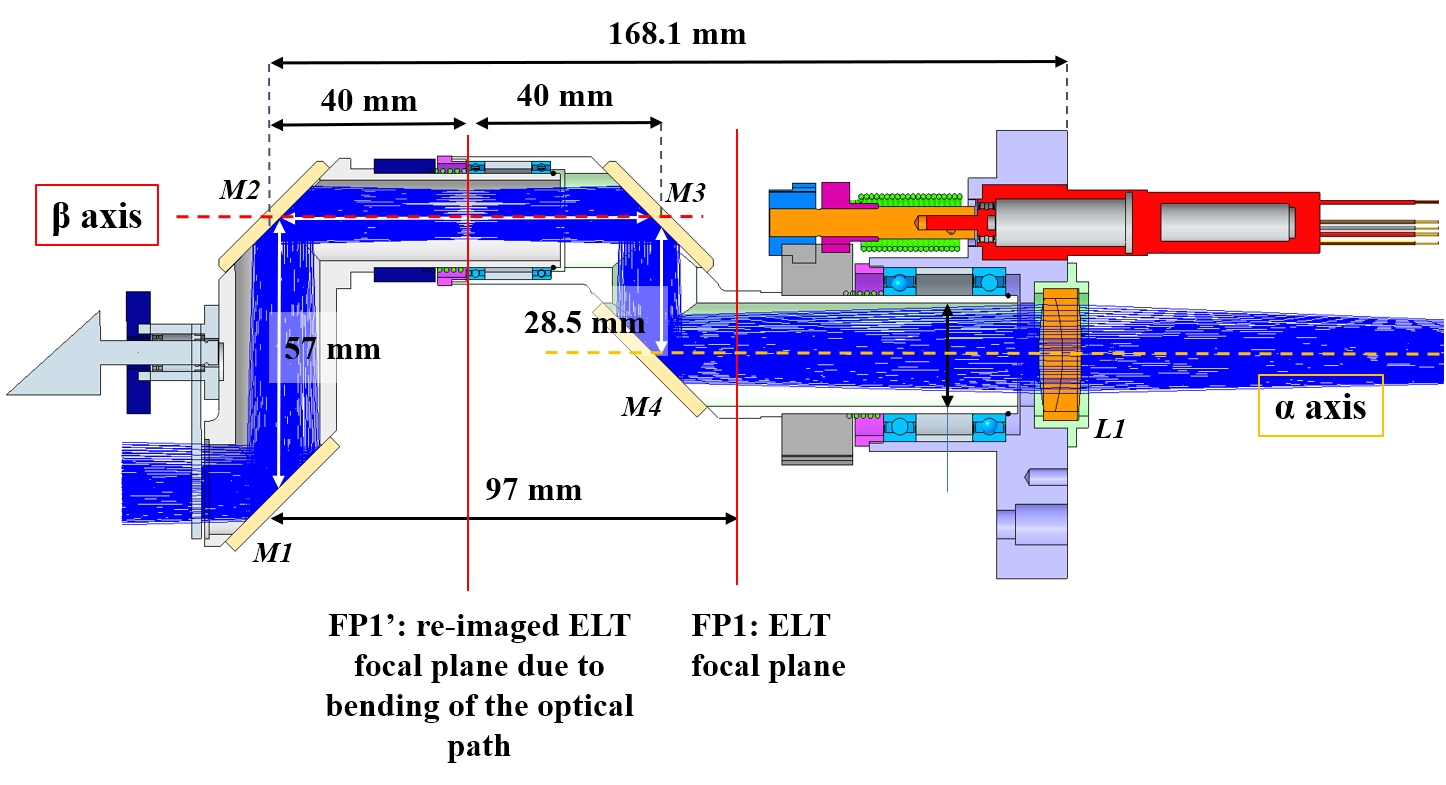}
         \caption{Cut view}
         \label{fig:pos_scara_meca_cut}
     \end{subfigure}
     \vspace{0.2cm}
        \caption{POS SCARA mechanics overview}
        \label{fig:pos_scara_meca}
\end{figure}

The theta-phi, or alpha-beta, architecture relies on two axes of rotation for the positioner to cover its donut-like patrol area. Figure \ref{fig:pos_scara_meca} highlights their actuation by two Brushless DC (BLDC) motors. This choice is built based on the known reliability of those motors, the expertise present in the EPFL Astrobots team on their control in position and their easiness to mass manufacture. The alpha and beta motor, respectively 13 and 8 mm, are coupled with Hall sensors, while the POM motor is coupled with an 1024 steps encoder due to the higher need in precision. We are using the ECXSP13M BL KL A STD 24V from Maxon coupled with the gearbox GPX16 A 1526:1 for the alpha motor and the other four motors in the MOSAIC POS are ECXSP08M BL KL A STD 6V configured with GPX08 A 1024:1 also from Maxon. This allows us to have only two motor references for the five motors of the MOSAIC POS.

\noindent Figure \ref{fig:pos_scara_meca_front} also shows, two backlash suppressor used to preload each rotary axis; removing the effect of backlash in the positioning routine. Together with the high gear ratio of both gearbox, they allow the positioner to be non back-drivable and hold the position while power is cut and the Instrument Core Structure (ICOS) is rotating, hence changing the gravity vector on the positioner; see Figure \ref{fig:icos} and \ref{fig:pos_modes}.

\begin{table}[H]
\centering
\caption{Mirror support faces nominal angles also referred as "telecentric angles"}
\label{tab:mirrors_faces_angles}
\begin{tabular}{@{}ll@{}}
\toprule
                    & Angle            \\ \midrule
M1 to $\beta$ axis  & 44.9570$\degree$  \\
M2 to $\beta$ axis  & 45$\degree$      \\
M3 to $\alpha$ axis & 44.9785$\degree$ \\
M4 to $\beta$ axis  & 45$\degree$      \\ \bottomrule
\end{tabular}
\end{table}

\subsection{POS ADC}
While the Atmospheric Dispersion Corrector (ADC) may be part of the instrument Front-End itself such as in DESI\cite{miller_optical_2024} or the telescope subsystem like VLT Survey Telescope, MOSAIC had to adapt, once again, for the unusual size of the ELT. Indeed, the optics for an ADC that may cover the full size of MOSAIC Focal Plane (more than 2 meters) was deemed to expensive in risks and resources. The size of the counter-rotating prisms in addition to the necessary mechanics to put them in motion would also have been outside the allocated space envelope on the Nasmyth platform.\\
Decision was therefore made to move the ADC at the positioners level, where the optics are down to a more reasonable size; the cylindrical prisms are 20.2 mm in diameter and 48 mm long.
\begin{figure}
    \centering
    \includegraphics[width=\linewidth]{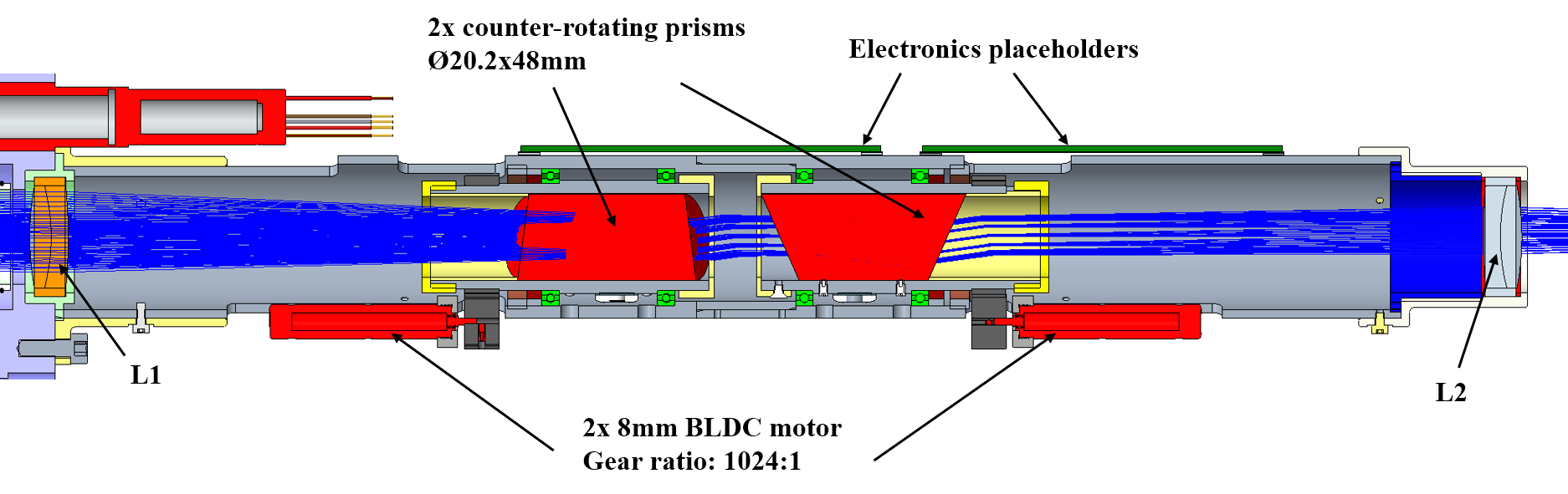}
    \vspace{0.2cm}
    \caption{Cut view of the ADC insides; highlighting the two counter-rotating prisms}
    \label{fig:ADC_cut}
\end{figure}

\subsection{POS interface with Focal Plane}

\begin{figure}[H]
    \centering
    \includegraphics[width=0.9\linewidth]{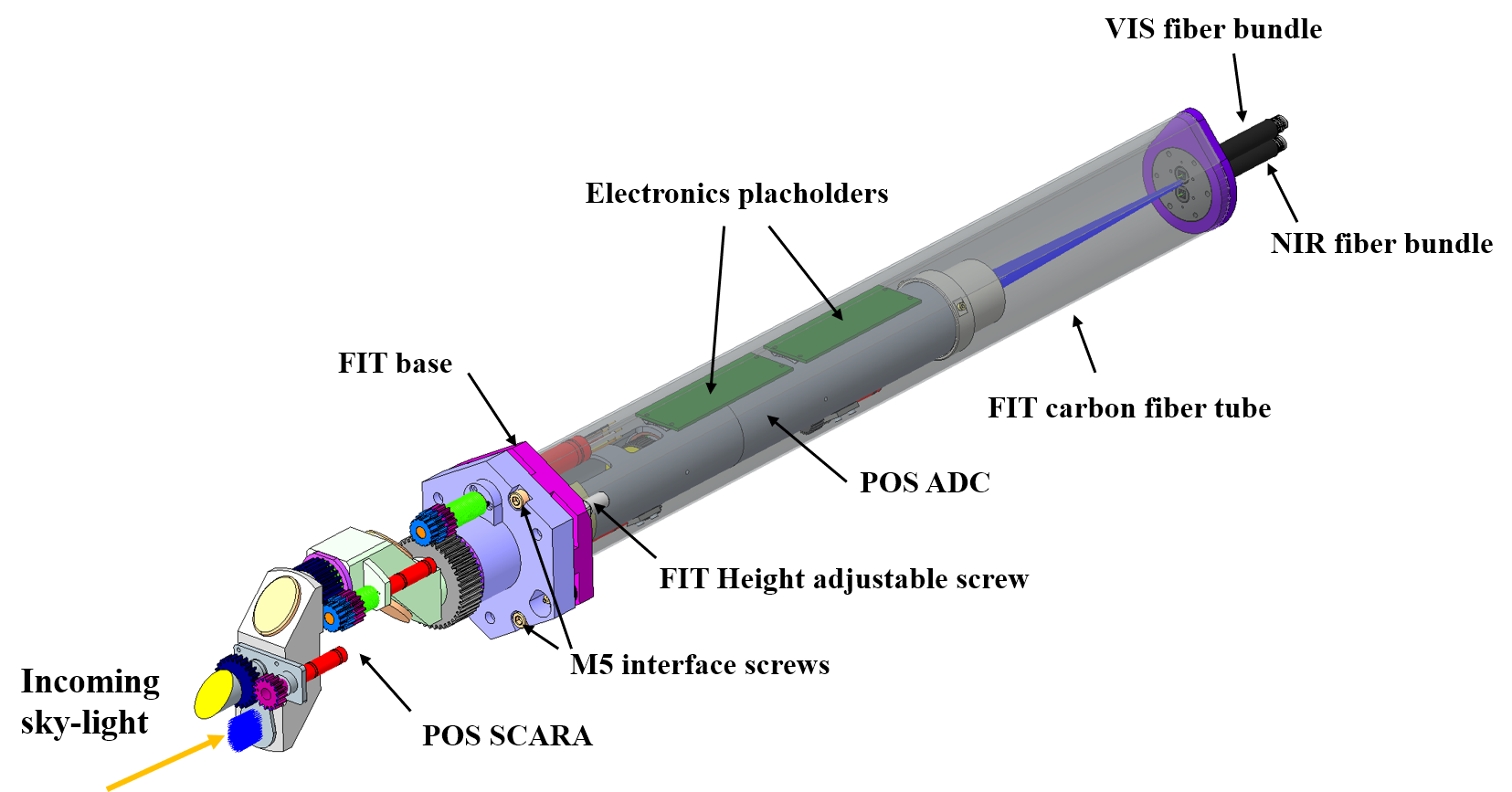}
    \vspace{0.2cm}
    \caption{POS inserted into FIT overview}
    \label{fig:POS_FIT_overview}
\end{figure}

Light entering the MOS-path needs to be focused on either a VIS or NIR fiber bundle located at the end of the optical path; see Figure \ref{fig:pos_modes}. For maintenance purposes we want to keep the ability to remove a positioner (SCARA and ADC) from the MOSAIC Focal Plane without needing to access the fiber bundles. Decision was therefore made to interface the positioner with the Focal Plane through a Fiber Interface Tube (FIT).\\
Under the responsibility of the Focal Plane subsystem, it consists of an aluminum base (in fuchsia in Figure \ref{fig:POS_FIT_overview}), a carbon fiber tube and an aluminum plate holding the fiber bundles.\\
POS and FIT have a flat interface and their respective base plate are positioned and bolted together by three M5 screws and 2 dowel pins, as seen in Figure \ref{fig:pos_scara_meca_back} and \ref{fig:POS_FIT_overview}. Their orientation in the Focal Plane is given by the 3 height adjustable screws at the back of the FIT base.\\

\begin{figure}[H]
    \centering
    \includegraphics[width=0.4\linewidth]{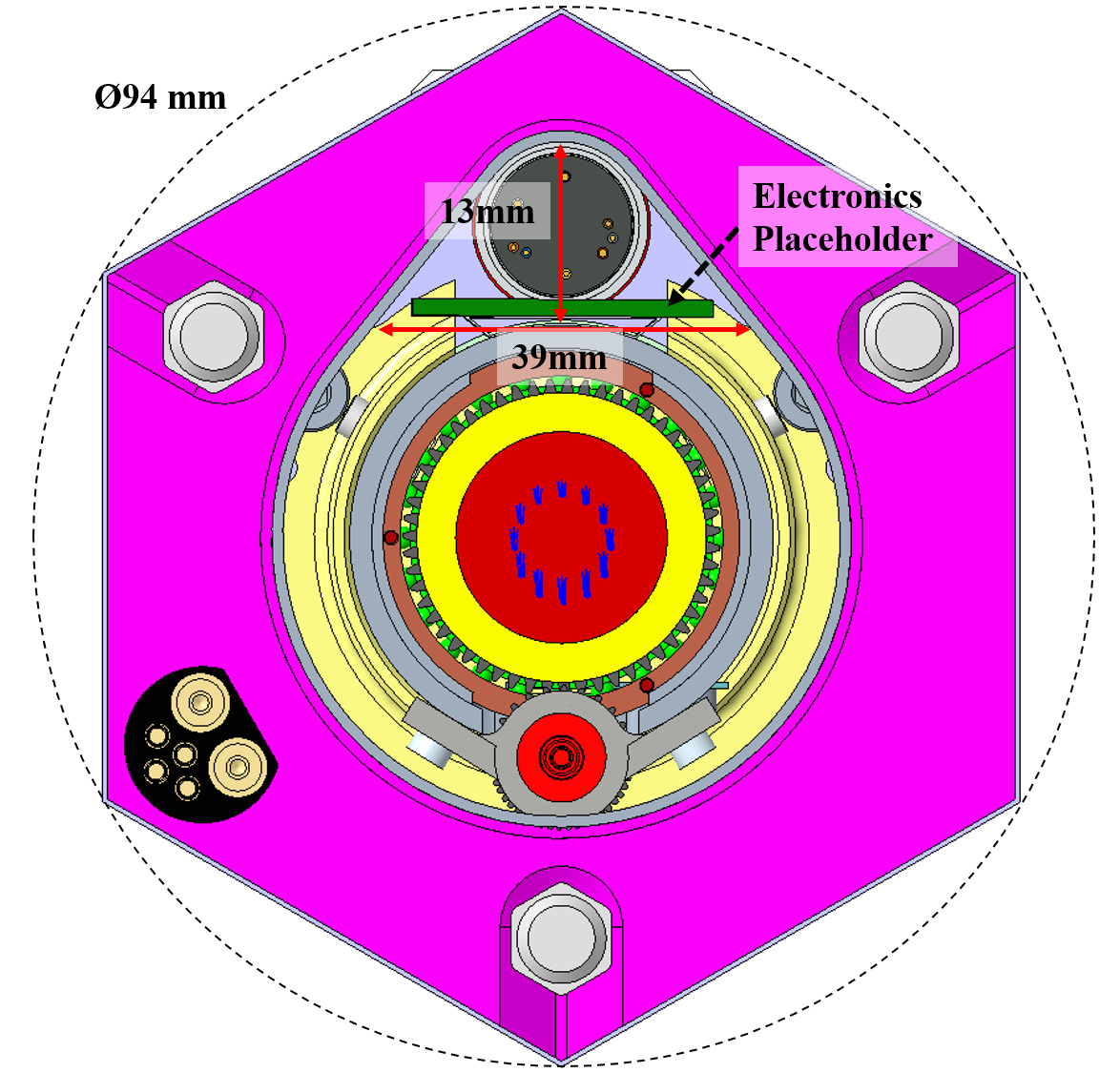}
    \vspace{0.2cm}
    \caption{Cut view from the back of the POS/FIT assembly}
    \label{fig:POS_FIT_cut_back}
\end{figure}

While the POS/FIT encapsulation improves the maintainability of MOSAIC, it limits the available space for the control electronics. Those are foreseen to be bolted on the ADC tube. The available space is therefore limited with the inner surface of the FIT: 13 mm height, 39 mm wide and 200 mm long, see Figure \ref{fig:POS_FIT_cut_back}.\\

\subsection{POS electronics}
\begin{figure}[H]
    \centering
    \includegraphics[width=0.6\linewidth]{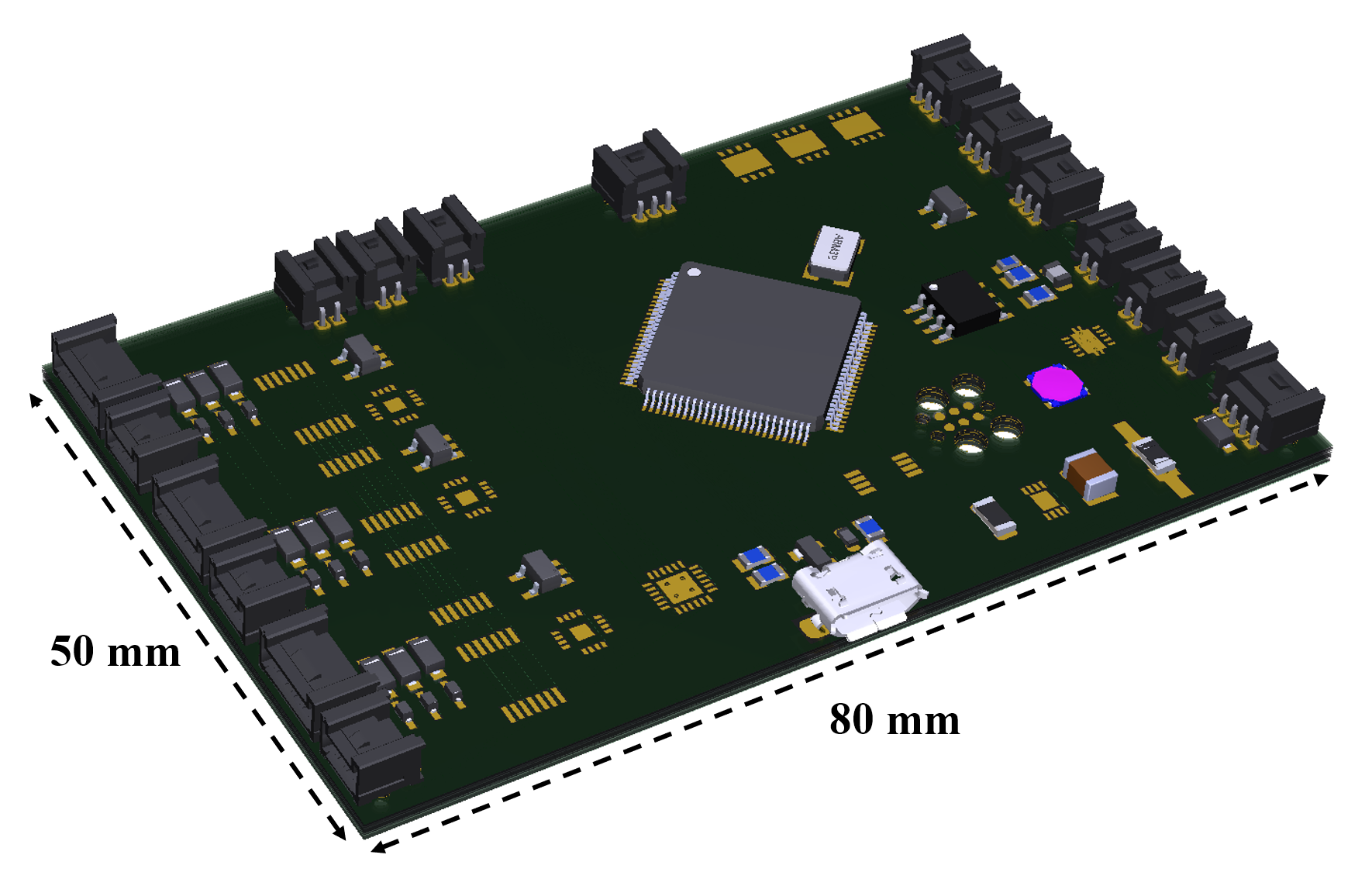}
    \caption{POS Control Board V1: 3D model}
    \label{fig:pos_electronics}
\end{figure}

Figure \ref{fig:POS_FIT_cut_back} highlights the small available space for control electronics in the MOSAIC positioner. As a reminder the current main electronics components to drive are:
\begin{itemize}
    \item 1x 13 mm BLDC motor 24V with Hall sensors
    \item 3x 8 mm BLDC motor 6V with Hall sensors
    \item 1x 8 mm BLDC motor 6V with rotary encoder
\end{itemize}
There is currently no off-the-shelf solution available to perform the control in position of those BLDC motors packed in such a tight space. A collaboration is in place between EPFL and USP to prototype the MOSAIC POS control board. The V1 design is portrayed in Figure \ref{fig:pos_electronics}. It will serve as a first test platform to try components and firmware. Its goal is not yet to fit in the before-mentioned tight space. This will be left to the V2.

\subsection{POS SCARA calibration}

The pointing of the positioner SCARA will be calibrated in the lab for each unit. For each fiber bundle, a look-up table will be characterized, linking the location of the star in the ELT focal plane FP1 to a specific configuration of the alpha and beta arms, for the ADC neutral wavelength. Each Positioner will be delivered to the Focal Plane sub-system calibrated, with its 2 look-up tables file. The test bench concept is described in Schmoll et al. (2026) \cite{schmoll_mosaic_2026}

\section{POS first prototyping}
\label{sec:pos_proto}
\begin{figure}[H]
     \centering
     \begin{subfigure}[b]{0.49\textwidth}
        \centering
        \includegraphics[width=\linewidth]{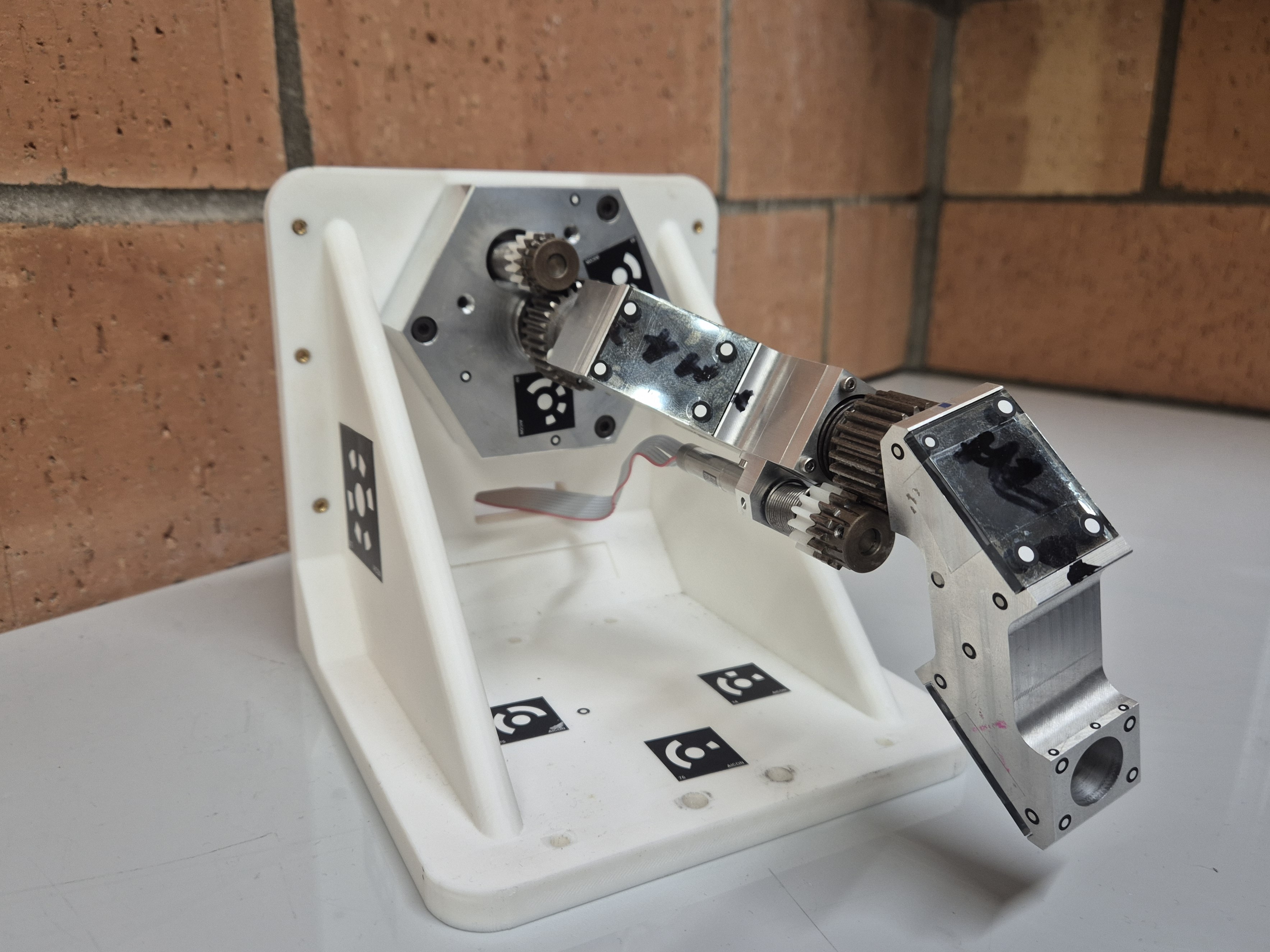}
        \caption{SCARA V1 prototype assembled}
        \label{fig:004_pos_proto_scara}
     \end{subfigure}
     \hfill
     \begin{subfigure}[b]{0.49\textwidth}
         \centering
         \includegraphics[width=\linewidth]{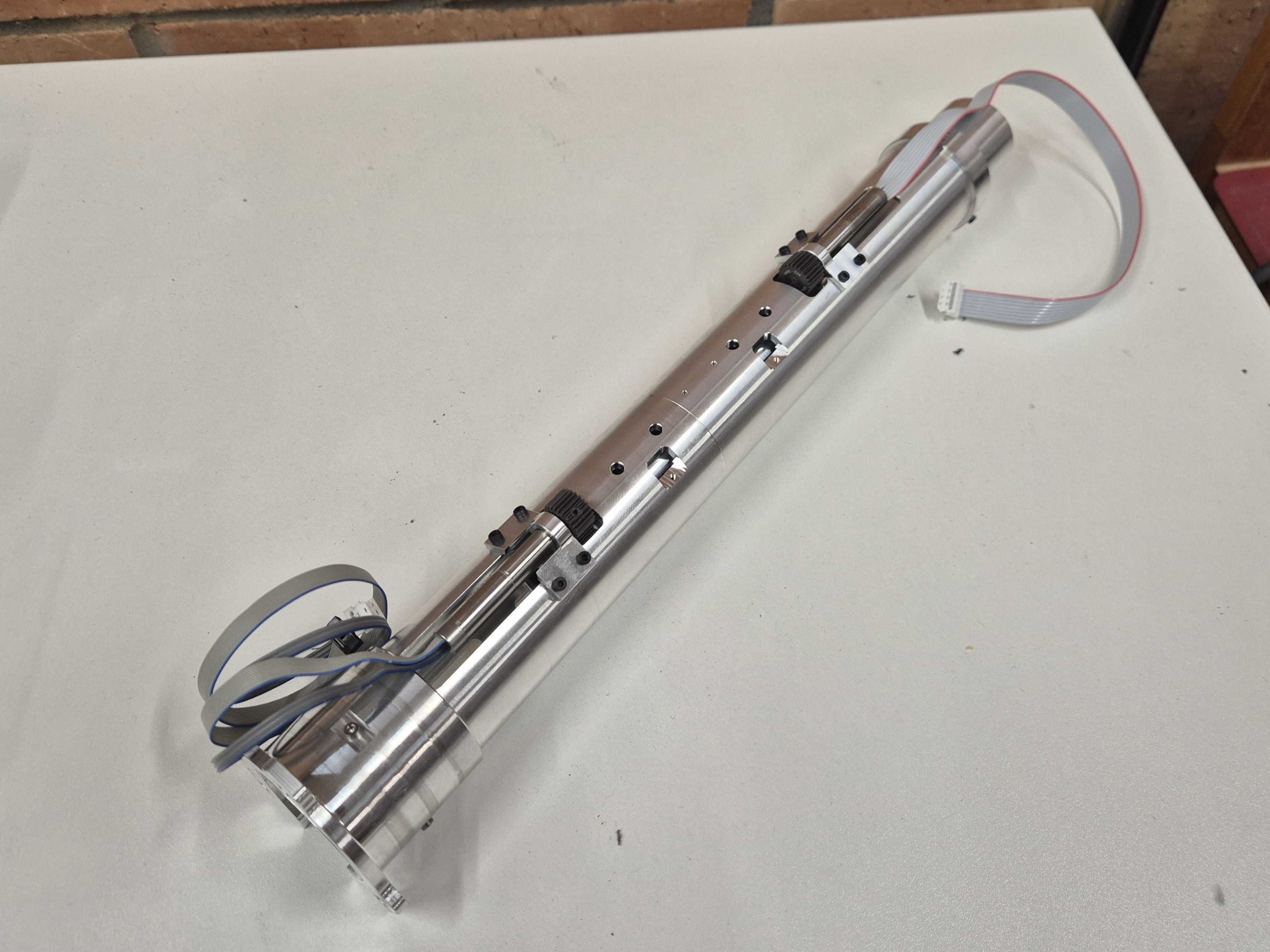}
         \caption{ADC V1 prototype assembled}
         \label{fig:004_pos_proto_adc}
     \end{subfigure}
     \vspace{0.2cm}
        \caption{MOSAIC POS V1 prototype}
        \label{fig:pos_proto_v1}
\end{figure}

The first stage of prototyping of the MOSAIC positioner aims to fabricate a SCARA V1 and an ADC V1 to validate mechanical manufacturability and assembly. Figure \ref{fig:pos_proto_v1} shows the two realized prototypes that are now being tested in EPFL.

\acknowledgments 
The authors acknowledges the mechanical workshop of the EPFL Insitute of Physics for their manufacturing of the ADC V1 and the workshop of HEPIA for the realization of the SCARA V1.\\
We also aknowledge the financial support of the Swiss National Science Fundation (SNSF) - Funding LArge international REsearch (FLARE) program that enables us to build this existing project.

\bibliography{references_FIXED}

@inproceedings{schallig_mosaic_2026,
	title = {{MOSAIC} at the {ELT}: focal plane architecture and initial design},
	volume = {14149},
	shorttitle = {},
	url = {},
	doi = {},
	language = {en},
	urldate = {},
	booktitle = {Ground-based and {Airborne} {Instrumentation} for {Astronomy} {XI}},
	publisher = {SPIE},
	author = {Ellen Schallig and et al.},
	month = jul,
	year = {2026},
	pages = {266 },
}

@misc{cvetojevic_mosaic_2026,
	title = {{MOSAIC} at {ELT}: the optical relay sub-system ({ORSS}) {\textbar} {SPIE} {Astronomical} {Telescopes} + {Instrumentation}},
	shorttitle = {{MOSAIC} at {ELT}},
	url = {https://spie.org/astronomical-telescopes-instrumentation/presentation/From-Manufacturing-to-Optical-Performance--Validation-of-the-GIRMOS/14149-286},
	language = {en},
	urldate = {2026-06-17},
	author = {Cvetojevic, Nick},
	year = {2026},
}

@article{miller_optical_2024,
	title = {The {Optical} {Corrector} for the {Dark} {Energy} {Spectroscopic} {Instrument}},
	volume = {168},
	issn = {0004-6256, 1538-3881},
	url = {https://iopscience.iop.org/article/10.3847/1538-3881/ad45fe},
	doi = {10.3847/1538-3881/ad45fe},
	language = {en},
	number = {2},
	urldate = {2026-06-01},
	journal = {The Astronomical Journal},
	author = {Miller, Timothy N. and Doel, Peter and Gutierrez, Gaston and Besuner, Robert and Brooks, David and Gallo, Giuseppe and Heetderks, Henry and Jelinsky, Patrick and Kent, Stephen M. and Lampton, Michael and Levi, Michael E. and Liang, Ming and Meisner, Aaron and Sholl, Michael J. and Silber, Joseph Harry and Sprayberry, David and Aguilar, Jessica Nicole and De La Macorra, Axel and Eisenstein, Daniel and Fanning, Kevin and Font-Ribera, Andreu and Gaztañaga, Enrique and Gontcho A Gontcho, Satya and Honscheid, Klaus and Jimenez, Jorge and Joyce, Dick and Kehoe, Robert and Kisner, Theodore and Kremin, Anthony and Landriau, Martin and Le Guillou, Laurent and Magneville, Christophe and Martini, Paul and Miquel, Ramon and Moustakas, John and Nie, Jundan and Percival, Will and Poppett, Claire and Prada, Francisco and Rossi, Graziano and Schlegel, David and Schubnell, Michael and Seo, Hee-Jong and Sharples, Ray and Tarlé, Gregory and Vargas-Magaña, Mariana and Zhou, Zhimin and {the DESI Collaboration}},
	month = aug,
	year = {2024},
	pages = {95},
}

@inproceedings{rodrigues_developing_2016,
	title = {Developing an integrated concept for the {E}-{ELT} {Multi}-{Object} {Spectrograph} ({MOSAIC}): design issues and trade-offs},
	shorttitle = {Developing an integrated concept for the {E}-{ELT} {Multi}-{Object} {Spectrograph} ({MOSAIC})},
	url = {http://arxiv.org/abs/1609.06610},
	doi = {10.1117/12.2232562},
	language = {en},
	urldate = {2026-05-29},
	author = {Rodrigues, Myriam and Dalton, Gavin and Fitzsimons, Ewan and Chemla, Fanny and Morris, Tim and Hammer, Francois and Puech, Mathieu and Evans, Christopher and Jagourel, Pascal},
	month = aug,
	year = {2016},
	note = {arXiv:1609.06610 [astro-ph.IM]},
	pages = {99089S},
}

@inproceedings{sayres_sdss-v_2022,
	title = {{SDSS}-{V} robotic focal plane system: overview of coordinate systems and transforms},
	volume = {12184},
	shorttitle = {{SDSS}-{V} robotic focal plane system},
	url = {https://www.spiedigitallibrary.org/conference-proceedings-of-spie/12184/121847K/SDSS-V-robotic-focal-plane-system--overview-of-coordinate/10.1117/12.2630507},
	doi = {10.1117/12.2630507},
	language = {en},
	urldate = {2026-05-29},
	booktitle = {Ground-based and {Airborne} {Instrumentation} for {Astronomy} {IX}},
	publisher = {SPIE},
	author = {Sayres, Conor and Sánchez-Gallego, José R. and Blanton, Michael R. and Engelman, Michael and Finkbeiner, Douglas P. and Hogg, David W. and Holtzman, Jon A. and Jurgenson, Colby and Pogge, Richard W. and Ramírez, Solange and Saydjari, Andrew K. and Schlafly, Edward F. and Tuttle, Sarah},
	month = aug,
	year = {2022},
	pages = {2412},
}

@inproceedings{schubnell_desi_2016,
	title = {The {DESI} fiber positioner system},
	volume = {9908},
	url = {https://www.spiedigitallibrary.org/conference-proceedings-of-spie/9908/990892/The-DESI-fiber-positioner-system/10.1117/12.2233370},
	doi = {10.1117/12.2233370},
	language = {en},
	urldate = {2026-05-29},
	booktitle = {Ground-based and {Airborne} {Instrumentation} for {Astronomy} {VI}},
	publisher = {SPIE},
	author = {Schubnell, Michael and Ameel, Jon and Besuner, Robert W. and Gershkovich, Irena and Hoerler, Philipp and Kneib, Jean-Paul and Heetderks, Henry D. and Silber, Joseph H. and Tarlé, Gregory and Weaverdyck, Curtis},
	month = aug,
	year = {2016},
	pages = {2715},
}

@misc{schmoll_mosaic_2026,
	title = {{MOSAIC}, the {ELT} multi-object spectrograph: {Positioners} with integrated atmospheric dispersion correction {\textbar} {SPIE} {Astronomical} {Telescopes} + {Instrumentation}},
	shorttitle = {{MOSAIC}, the {ELT} multi-object spectrograph},
	url = {https://spie.org/astronomical-telescopes-instrumentation/presentation/MOSAIC-the-ELT-multi-object-spectrograph--Positioners-with-integrated/14149-282},
	language = {en},
	urldate = {2026-05-08},
	author = {Schmoll, Jurgen},
	year = {2026},
}

@inproceedings{morris_eagle_2012,
	address = {Amsterdam, Netherlands},
	title = {The {EAGLE} instrument for the {E}-{ELT}: developments since delivery of {Phase} {A}},
	shorttitle = {The {EAGLE} instrument for the {E}-{ELT}},
	url = {http://proceedings.spiedigitallibrary.org/proceeding.aspx?doi=10.1117/12.925889},
	doi = {10.1117/12.925889},
	language = {en},
	urldate = {2026-05-08},
	author = {Morris, Simon L. and Cuby, Jean-Gabriel and Dubbeldam, Marc and Evans, Christopher and Fusco, Thierry and Jagourel, Pascal and Myers, Richard M. and Parr-Burman, Phil and Rousset, Gerard and Schnetler, Hermine},
	editor = {McLean, Ian S. and Ramsay, Suzanne K. and Takami, Hideki},
	month = sep,
	year = {2012},
	pages = {84461J},
}

@article{hammer_optimos-eve_2010,
	title = {{OPTIMOS}-{EVE}: {A} {Fibre}-fed {Optical}-{Near}-infrared {Multi}-object {Spectrograph} for the {E}-{ELT}},
	volume = {140},
	issn = {0722-6691},
	shorttitle = {{OPTIMOS}-{EVE}},
	url = {https://ui.adsabs.harvard.edu/abs/2010Msngr.140...36H},
	urldate = {2026-05-08},
	journal = {The Messenger},
	author = {Hammer, F. and Kaper, L. and Dalton, G.},
	month = jun,
	year = {2010},
	note = {ADS Bibcode: 2010Msngr.140...36H},
	pages = {36--37},
}

@inproceedings{gonzalez_moons_2022,
	title = {{MOONS} – multi object spectroscopy for the {VLT}: overview and instrument integration update},
	volume = {12184},
	shorttitle = {{MOONS} – multi object spectroscopy for the {VLT}},
	url = {https://www.spiedigitallibrary.org/conference-proceedings-of-spie/12184/1218412/MOONS--multi-object-spectroscopy-for-the-VLT--overview/10.1117/12.2629932.full},
	doi = {10.1117/12.2629932},
	urldate = {2026-05-08},
	booktitle = {Ground-based and {Airborne} {Instrumentation} for {Astronomy} {IX}},
	publisher = {SPIE},
	author = {Gonzalez, Oscar and et al.},
	month = aug,
	year = {2022},
	pages = {342--357},
}

\bibliographystyle{spiebib} 


\end{document}